\documentclass[twocolumn,british,aps,prl,superscriptaddress,showpacs]
{revtex4-1}

\usepackage{etoolbox}
\newbool{arxive}

\booltrue{arxive}  

\ifbool{arxive}{
\pdfoutput=1
\usepackage[final]{pdfpages}
}{}

\usepackage[T1]{fontenc}
\usepackage[latin1]{inputenc}
\usepackage{amsmath}

\usepackage{graphicx}
\makeatletter

\usepackage{babel}
\makeatother

\newcommand{\bpsi}{\mbox{\boldmath $\psi$}}
\newcommand{\nnabla}{\mathbf \nabla}

\newcommand{\req}[1]{Eq.~(\ref{#1})}
\newcommand{\reqs}[1]{Eqs.~(\ref{#1})}
\newcommand{\rref}[1]{(\ref{#1})}

\renewcommand{\P}{\mathbf{P}}
\newcommand{\n}{\mathbf{n}}
\renewcommand{\r}{\mathbf{r}}

\newcommand{\beq}{\begin{equation}}
\newcommand{\eeq}{\end{equation}}
\newcommand{\be}{\begin{equation}}
\newcommand{\ee}{\end{equation}}
\newcommand{\beqa}{\begin{eqnarray}}
\newcommand{\eeqa}{\end{eqnarray}}
\newcommand{\bea}{\begin{eqnarray}}
\newcommand{\eea}{\end{eqnarray}}
\usepackage{amssymb}
\usepackage{array}
\usepackage{amsmath}
\usepackage{graphicx}\usepackage{graphics}
\usepackage{dcolumn}
\usepackage{bm}\usepackage{varioref}

\newcommand{\hLambda}{\hat{\Lambda}}

\newcommand{\R}{\mathbf R}

\begin{document}

\title{Three-Particle Complexes in Two-Dimensional Semiconductors}

\author{Bogdan Ganchev}
\affiliation{Department of Physics, Lancaster University, Lancaster,
  LA1~4YB, United Kingdom}

\author{Neil Drummond}
\affiliation{Department of Physics, Lancaster University, Lancaster,
  LA1~4YB, United Kingdom}

\author{Igor Aleiner}
\affiliation{Physics Department, Columbia University, New York, NY 10027, USA}
\affiliation{Department of Physics, Lancaster University, Lancaster,
  LA1~4YB, United Kingdom}

\author{Vladimir Fal'ko}
\affiliation{Department of Physics, Lancaster University, Lancaster,
  LA1~4YB, United Kingdom}

\begin{abstract}
We evaluate binding energies of trions $X^{\pm}$, excitons bound by a donor/acceptor charge $X^{D/A}$, 
and overcharged acceptors/donors in two-dimensional atomic crystals by mapping the three-body problem in two dimensions 
onto one particle in a three dimensional potential treatable by a purposely-developed boundary-matching-matrix method. 
We find that in monolayers of transition metal dichalcogenides the dissociation energy of $X^{\pm}$ is typically much larger than that of localised exciton complexes, so that trions are more resilient to heating, despite the fact that their recombination line in optics is less red-shifted from the exciton line than the line of $X^{D/A}$. 
\end{abstract}

\date{\today}
\maketitle

Atomic layers of hexagonal transition metal dichalcogenides (TMDC) \cite{TMDC1,TMDC2,TMDC3,TMDC4}, represent a new
class of systems whose optical properties attract a lot of interest \cite{2Doptical1,2Doptical2,2Doptical3,2Doptical4,2Doptical5,2Doptical6,TMDC4}, due to their promise for applications in optoelectronics.  
These two-dimensional (2D) crystals are believed to be direct band gap semiconductors 
\cite{2DBands1,2DBands2,2DBands3,2DBands4,2DBands5}, and their
luminescence spectra contain distinct lines interpreted as the electron-hole recombination from neutral, $X$, and charged excitons
(trions $X^\pm$) \cite{Excitons1,Excitons2,Excitons3,Excitons4,Excitons5,Excitons6,Trions1,Trions2,Trions3,TMDC4}, which also coexist 
with the recombination of excitons localised at defects.

Here, we study binding energies of acceptor/donor-bound excitons ($X^{A/D}$),  
trions ($X^\pm$) and charged acceptors/donors ($A^+/D^-$) in atomic 2D 
crystals using a method developed specifically to tackle such three-body problems in two dimensions. 
For the trions, we also employ the diffusion quantum Monte Carlo approach \cite{Ceperley_1980,Needs_2010}. 
We take into account a specific feature of atomically thin crystals of TMDCs, 
where, due to the polarisability of atomic orbitals, the interaction between charges $q_{i,j}$ is logarithmic, 
$\frac{q_iq_j}{r_*}\ln\frac{r_{ij}}{r_*}$,
up to a distance $r_*$ much larger than the excitonic Bohr radius \cite{footnote}, 
as inducated by the comparison of measured 
\cite{Heinz} and calculated \cite{Excitonlog1,Excitonlog2,Heinz} spectra of ground and excited states of free excitons. 

\begin{figure}[h]
\includegraphics[width=0.9\columnwidth]{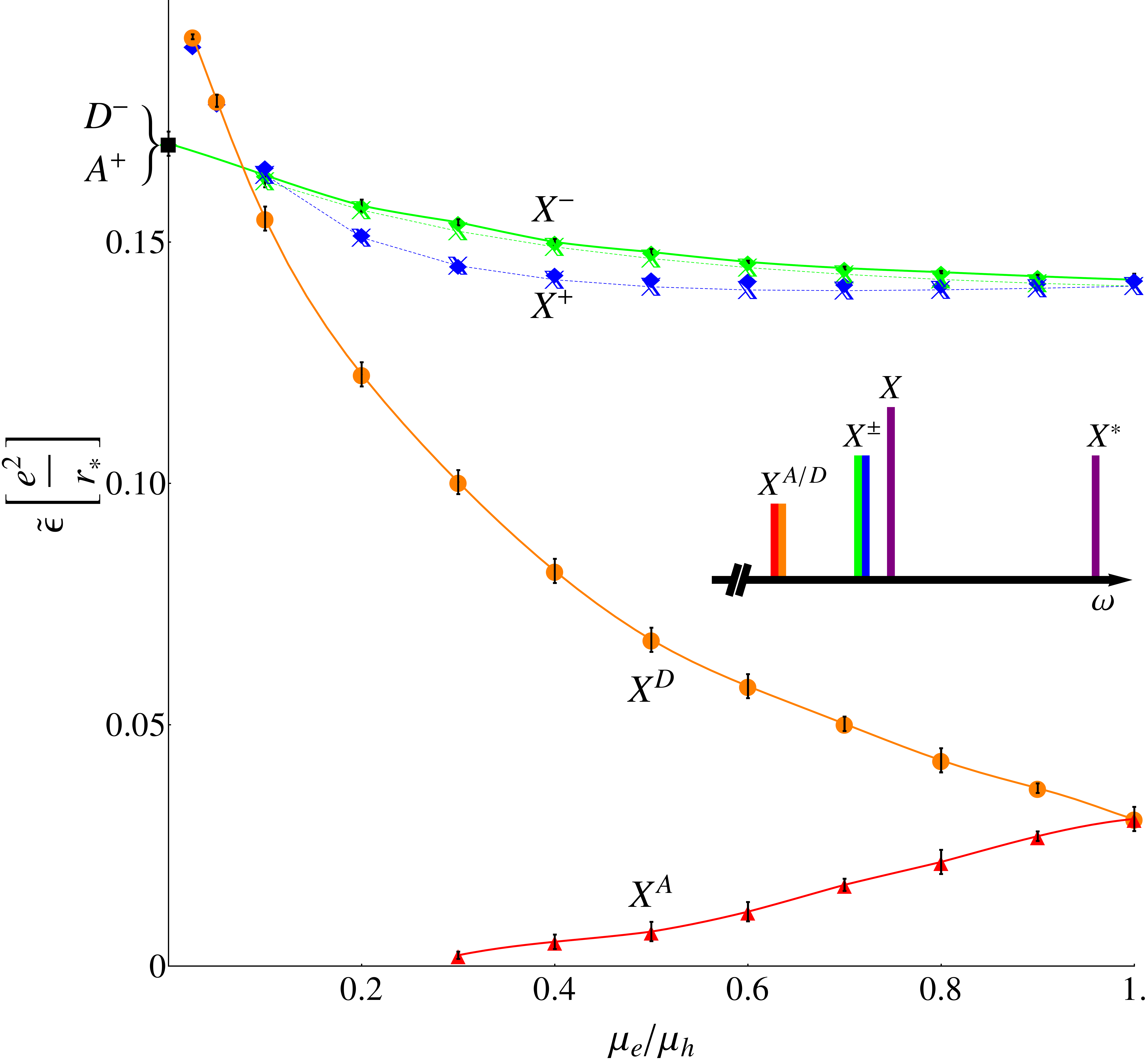}
\caption{Binding energies, $\tilde{\epsilon}$ of charged complexes $X^{A/D}$, $X^\pm$, $A^+/D^-$ for various electron-hole mass ratios, $\mu_e/\mu_h<1$ 
(for MoS$_2$, and MoSe$_2$, $\mu_e/\mu_h\approx0.7$; for WS$_2$ and WSe$_2$, $\mu_e/\mu_h\approx0.6$, \cite{2DBands1,2DBands5,masses1,masses2,masses3,masses4}). 
For trions, the results obtained by the newly developed method (diamonds) are compared to the binding energies determined using the diffusion Monte Carlo technique (crosses).
{\it Sketch}: sequence of luminescence lines in TMDC spectra, including charged complexes as well as ground and first radiative excited states of the free exciton. 
}
\label{fig3}
\end{figure}

In Fig.~\ref{fig3} we display the calculated binding energies $\tilde{\epsilon}$ of all charged three-particle complexes 
which determine the activation energy needed to dissociate them into a neutral complex and a free carrier ($X^{\pm}\rightarrow X+e/h$; $X^{D/A}\rightarrow D^0/A^0+h/e$). 
For the parametric range $0.5<\frac{\mu_e}{\mu_h}<2$, which covers the effective masses of MoS$_{2}$, WS$_{2}$, MoSe$_{2}$ and WSe$_{2}$ \cite{2DBands7},
we find that the discociation of $X^{D/A}$ into a neutral donor/acceptor and a hole/electron 
has a much smaller activation threshold than the dissociation of a trion, which suggests that in TMDC luminescence the stronger 
red-shifted $X^{A/D}$ line would be more sensitive to temperature than the trion line.

\begin{figure}[h]
\includegraphics[width=0.7\columnwidth]{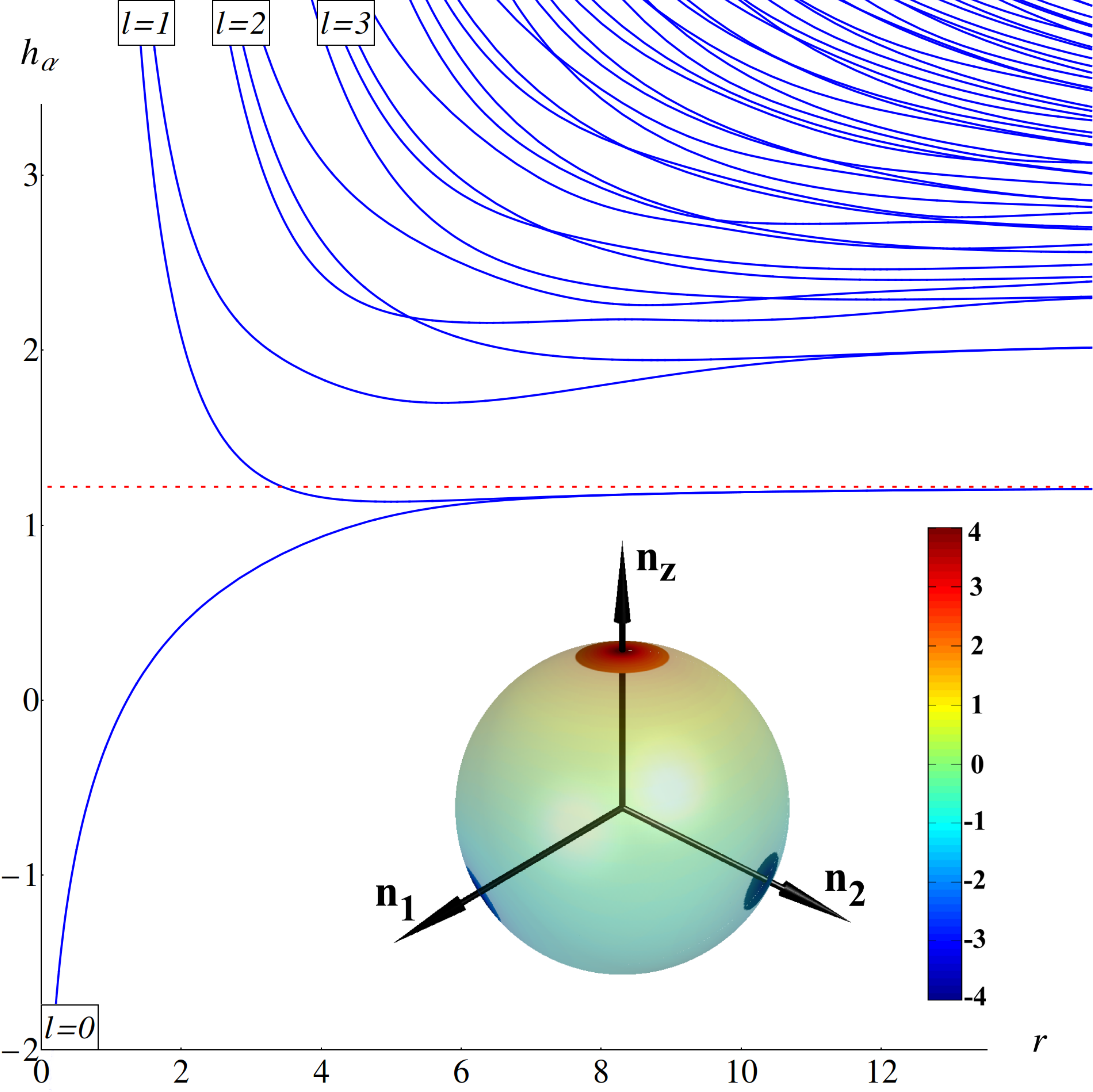}
\caption{ 
 Eigenvalues $h_{\alpha}(r)$ of $\hat{H}(r)$ in \req{matrixelements} for $\theta_1=\theta_2=\frac{\pi}{2}$ and $L_{max}=30$
   \cite{FootnoteFig}.  For $r\ll 1$,  $h_{\alpha}$
   are bunched by the angular momenta $l$, whereas for $r\gg 1$, $a/s$ doublets
   correspond to the particle localization in the minima $\n_{1,2}$ with vanishing tunneling (note that $a-s$ crossings are allowed).  
   The red dashed line marks the boundary $X_0^0$ of the continuum spectrum for the exciton
   and a free particle, and at $r\gg 1$, $h(r) \approx X_0^0 - c/r^2$, 
   determined by the 2D van der Waals attraction 
   between the charged particle and the neutral exciton, which produces an infinite number of shallow bound states.
{\em Inset:} Color scale image of the potential $U$ in
   \req{problem3}.  
}
\label{fig1}
\end{figure}

Since most of the results displayed in Fig.~\ref{fig3} were obtained using an original approach, 
we describe its logic and theoretical features in detail, whereas the diffusion quantum Monte Carlo calculations \cite{Ceperley_1980,Needs_2010} 
are discussed in Supplementary Information (SI) \cite{Sup}. 
Three 2D particles have six degrees of freedom, three of which correspond to centre-of-mass motion and overall rotation. 
The quantum mechanics of the remaining degrees of freedom is equivalent to that of a particle moving in an effective three-dimensional potential. 
The wave function  $\Psi(\r_1,\r_2,\r_3)$ of three logarithmically interacting particles with masses $\mu_{1,2,3}$ 
\cite{masses1,masses2,masses3,masses4,FootnoteParabolic} and charges $|q_i|=e$, 
$q_1q_2=e^2;\ q_3q_{1,2}=-e^2$, 
obeys the Schr\"odinger equation ($\r_{ij}\equiv \r_i-\r_j$),
\be
\begin{split}
&\left[
-\sum_{i=1}^3\frac{\hbar^2\nnabla^2_{\r_i}}{2\mu_i}
+ \frac{e^2}{r_*}\ln\frac{|\r_{31}||\r_{32}|}{r_*|\r_{12}|}
\right]\Psi=E\Psi.
\end{split}
\label{problem}
\ee
After separating the motion of the center of mass, 
$\R_{cm}=(\sum_i\mu_i\r_i)/M,\ \ M\equiv\sum_i\mu_i$, introducing dimensionless 
$\tilde{\r}=\r_{12}/{r_0};
 \  {\r}^\prime=\left(\sum_{i=1}^2\mu_i\r_{i3}\right)/\left(r_0\left[M \mu_1\mu_2/\mu_3\right]^{1/2}\right)$,
where $r_0^{-1}=[2e^2\mu_1\mu_2/
(r_*\hbar^2(\mu_1+\mu_2))]^{1/2}$, and spherical coordinates, 
\[
\begin{bmatrix} 
\left[r_x^\prime, r_y^\prime\right] \\[5pt] \left[\tilde{r}_x, \tilde{r}_y\right]\end{bmatrix}=r\begin{bmatrix}\cos\frac{\theta}{2}
\left[ \cos\left(\Phi+\frac{\phi}{2}\right),\
\sin\left(\Phi+\frac{\phi}{2}\right)\right]
\\[5pt]
\sin\frac{\theta}{2}\left[\cos\left(\Phi-\frac{\phi}{2}\right),\
\sin\left(\Phi-\frac{\phi}{2}\right)\right] \end{bmatrix},
\] 
\req{problem} takes the form
\be
\begin{split}
& E=\frac{\P_{cm}^2}{2M}+\frac{e^2}{r_*}
\left[\frac{1}{2}\ln\frac{\hbar^2\prod_{i=1,2}(\mu_i+\mu_3)}{2e^2r_*\mu_3^2(\mu_1+\mu_2)}
+\epsilon\right],
\end{split}
\ee
where $\epsilon$ are the eigenvalues of  Schr\"odinger
equation 
\be
\begin{split}
&\left[-\nnabla_4^2+\ln r + U(\theta,\phi)\right]\psi=\epsilon\psi;
\\
& -\nnabla_4^2 =-\frac{\partial^2}{\partial r^2}-\frac{3\partial}{r\partial r}
+ \frac{4\hat{\mathbf L}^2}{r^2} + \frac{\hat{\Theta}}{r^2\sin^2\theta };
\\
&U(\theta,\phi)=\frac{1}{2}\ln\left[\frac{(1-\n\cdot\n_1)(1-\n\cdot\n_2)}
{(1-\n\cdot\n_z)}\right],
\\
&\n= \left[\sin\theta\cos\phi,\ \sin\theta\sin\phi, \ \cos\theta\right].
\end{split}
\raisetag{30pt}
\label{problem3}
\ee 
This transforms
the three-body problem to a one-body problem in a higher-dimensional space, where $\n$ is a position of 
a fictitious particle on a sphere, ${\mathbf L}$ is a 3-dimensional angular momentum operator, 
$\hat{\mathbf L}^2=-
\frac{1}{\sin\theta}
\frac{\partial }{\partial \theta}\sin\theta\frac{\partial }{\partial \theta}- \frac{1}{\sin^2\theta}\frac{\partial^2}{\partial \phi^2}$,
$\hat{\Theta}=\left[-\frac{\partial}{\partial \Phi}+4\cos\theta\frac{\partial}{\partial \phi}\right]\frac{\partial}{\partial \Phi} $. 
Vectors  
$\n_z$ and $\n_{i=1,2}$ characterize the direction of the maximal repulsion and attraction, 
respectively,
\be
\begin{split}
& \n_z=\left[0,\ 0,\ 1\right]; \qquad
\n_i= \left[(-1)^i \sin\theta_i,\ 0, \ \cos\theta_i\right],
\\
& \tan\left({\theta_{1,2}}/{2}\right)=\left[
M\mu_{1,2}/({\mu_3\mu_{2,1}})
\right]^{1/2},
\end{split}
\label{angles}
\ee
where parameters for particular complexes are specified in Table~\ref{table1}. The color-scale visualization of $U$ is shown
in the inset to Fig.~\ref{fig1}. 
Classically, the particle collapses to either $\n_1$ or $\n_2$; 
  this observation is useful for finding the large-distance asymptotic states.

\begin{table*}[t]
\renewcommand{\arraystretch}{2.1}
\begin{tabular}{||c|c|c|c|c|c|c||}
\hline\hline
 &  \multicolumn{2}{c|}{$\parbox{5cm}{
Exciton ($X$) \\ localized on charged impurity}$} &  \multicolumn{2}{c|}{$\parbox{5cm}{Two particles \\ localized on charged impurity}$}  
& \multicolumn{2}{c||}{Trion}
\\
\cline{2-7}
& {Acceptor ($X^A$)} & Donor ($X^D$) & {Acceptor ($A^+$)} & Donor ($D^-$) & 
{Negative, $(X^-$)} & {Positive, $(X^+)$}
\\
\hline
  $\mu_1$ & $\mu_e$ &   $\mu_h$ & 
$\mu_h$ & $\mu_e$ & 
$\mu_e$ 
& $\mu_h$ 
\\
\hline
  $\mu_2$ & $\infty$ &   $\infty $ & 
$\mu_h$ & $\mu_e$ & 
$\mu_e$ 
& $\mu_h$ 
\\
\hline
  $\mu_3$ & $\mu_h$ &   $\mu_e$ & 
$\infty$ & $\infty $ & 
$\mu_h$ 
& $\mu_e$ 
\\
 \hline
  $\theta_1$ & $2\arctan \sqrt{\frac{\mu_e}{\mu_h}}$ & $2\arctan\sqrt{ \frac{\mu_h}{\mu_e}}$   & 
$\frac{\pi}{2}$ & $\frac{\pi}{2} $ & 
$2\arctan\sqrt{ \frac{2\mu_e+\mu_h}{\mu_h}}$  
& $2\arctan\sqrt{ \frac{2\mu_h+\mu_e}{\mu_e}}$ 
\\
 \hline
  $\theta_2$ & $\pi$ & $\pi$   & 
$\frac{\pi}{2}$ & $\frac{\pi}{2} $ & 
$2\arctan\sqrt{ \frac{2\mu_e+\mu_h}{\mu_h}}$  
& $2\arctan\sqrt{ \frac{2\mu_h+\mu_e}{\mu_e}}$  
\\
\hline\hline
\end{tabular}
\caption{Parameters in \reqs{problem}-\rref{angles} for charged complexes in 2D semiconductors with effective electron(hole) masses 
$\mu_{e(h)}$}.
\label{table1}
\end{table*}


Because of rotational symmetry, the potential $U$ in \req{problem3} 
does not depend on the angle $\Phi$. Hence, the eigenstates can be classified by the integer angular momentum 
$J$: $\Psi_J(r,\theta,\phi,\Phi)=e^{iJ\Phi}\psi_J(r,\theta,\phi)$, with $J=0$ for the ground state of the 3-particle complex (see SI \cite{Sup} for classification of $J\ne 0$). 
For  $\Psi_0$ to be single-valued, we must have $\psi_0(r,\theta+2\pi,\phi)=\psi_0(r,\theta,\phi+2\pi)=\psi_0(r,-\theta,\phi+\pi)=\psi_0(r,\theta,\phi)$. 
In general, the potential $U$ also has a mirror reflection symmetry $U(\phi)=U(-\phi)$. When two particles in the complex are identical ($\theta_1=\theta_2$),  
$U(\theta)=U(-\theta)$, states are either symmetric or anti-symmetric ($s/a$) in $\theta$.

In the following, we use the conventional \cite{Edmonds} basis of spherical harmonics $Y_{l\geq 0,m}(\theta,\phi),\ |m|\leq l$, 
\be
\begin{split}
&
\psi_{J=0}^e=\sum_{l=0}^{L_{max}}\sum_{m=0}^{l} Y_{l,m}^e(\n)\psi_{l,m}(r), \ \
\hat{\mathbf L}^2Y_{l,m}^e=l(l+1)Y_{l,m}^e;\\
&
Y^e_{l,0}=Y_{l,0};\ Y^e_{l,1 \leq m\leq l}=\frac{1}{\sqrt{2}}\left[Y_{lm}+(-1)^mY_{l,-m}\right],\\
&
Y_{l,m}(\theta,\phi)=(-1)^m\left[Y_{l,-m}(\theta,\phi)\right]^*=
(-1)^mY_{l,-m}(\theta,-\phi).
\end{split}
\nonumber
\raisetag{22pt}
\label{basis}
\ee

In this basis, \req{problem3} becomes 
\be
\left[\frac{d^2}{dr^2}+3\frac{d}{rdr}\right]\bpsi= 
\left(\hat{H}(r)-\epsilon{\openone}\right)\bpsi, 
\label{problemchannel}
\ee
where vector $\bpsi$ is comprised of the components $\psi_{l,m}$, and the matrix $\hat{H}(r)$ has elements 
\begin{widetext}
\be
\begin{split}
&H_{l_2m_2}^{l_1m_1}=
\left[
\frac{4l_1(l_1+1)}{r^2}+\frac{1}{2}\ln \frac{2r^2}{e}
\right]\delta_{l_1l_2}\delta_{m_1m_2}
+V_{l_2m_2}^{l_1m_1}, \
V_{l_2m_2}^{l_1m_1}\!\!=
\left[U_{l_2m_2}^{l_1m_1}+(-1)^{m_1}U_{l_2m_2}^{l_1,-m_1}\right]
\!\times\!\left\{\begin{matrix}
1, & m_{1,2} > 0;
\\
\frac{1}{2}, & m_{1,2}=0; \\
\frac{1}{\sqrt{2}}, & \text{otherwise}.
\end{matrix}\right.
\end{split}
\label{matrixelements}
\ee
Remarkably, the matrix elements $U_{l_2m_2}^{l_1m_1}$ can be found in a closed analytic form (for the derivation, see SI \cite{Sup}), 
\[
\begin{split}
U_{l_2m_2}^{l_1m_1}&=
(-1)^{m_1} \sqrt{\pi(2l_1+1)(2l_2+1)} 
\sum_{ l=l_{min}} ^{l_1+l_2}\frac{\sqrt{2l+1}}{l(l+1)}
\begin{pmatrix}l_1&l&l_2\\-m_1&m&m_2\end{pmatrix}
\begin{pmatrix}l_1&l&l_2\\0&0&0\end{pmatrix}
\left[Y^*_{l,m}(\n_z)-Y^*_{l,m}(\n_1)-Y^*_{l,m}(\n_2)\right],
\end{split}
\raisetag{45pt}
\]
\end{widetext}
where $m\equiv m_1-m_2$, $l_{min}\equiv {\rm max}(1,|l_1-l_2|,|m|)$, 
the $3j$ symbols follow  
 Ref.~\onlinecite{Edmonds}, and $\n_i$ are from \req{angles}.

Numerically found eigenvalues of
 Hamiltonian \rref{matrixelements} are shown in Fig.~\ref{fig1}. 
At $r\gg 1$, the eigenfunctions  
are peaked near $\n=\n_{1,2}$ suggesting an adiabatic solution 
for \req{problemchannel} at $r\gg 1$.
Consider the equation,
\be
\left[ -\frac{1}{\rho}\frac{d}{d\rho}\rho\frac{d}{d\rho} + \frac{m^2}{\rho^2}+ \ln \rho  \right] 
\varphi_n^m(\rho) = \chi_n^m\varphi_n^m(\rho), 
\label{exciton1}
\ee
which determines the spectrum 
of a 2D exciton
with the logarithmic interaction $e^2/r_*\ln(\rho/r_*)$: 
\be
X^m_n=\frac{e^2}{r^*}\left[\frac{1}{2}\ln\frac{\hbar^2(\mu_1+\mu_3)}{2e^2r_*\mu_1\mu_3}+\chi_n^m\right].
\label{exciton2}
\ee
Integer $m$ and $n\geq 0$ are the 2D angular momentum and radial quantum number, respectively, and 
the interlevel distances \cite{Heinz,Excitonlog2} determined by the eigenvalues listed in Table~\ref{table2} 
do not depend on the masses.  

\begin{table}[h]
\begin{tabular}{c||c|c|c}
$\chi_n^m$ & $m=0$ & $m=1$ & $m= 2$\\[0.1cm] 
\hline\hline\ 
$n=0$\ & \ $0.5265$ \ &\ $1.386$ \ & $1.844$\\[0.1cm]
\hline\
$n=1$\ & \ $1.661$ \ &\ $2.009$ \ & \\[0.1cm]
\hline\
$n=2$\ & \ $2.177$ \ &\  \ & 
\\[0.1cm]
\hline\hline
\end{tabular}
\caption{The eigenvalues of \req{exciton1} which determine the spectrum of ground and excited states of the exciton, \req{exciton2}.}
\label{table2}
\end{table}

The adiabatic 
wave function (closely bound electron-hole pair and the third particle far from the pair) 
is 
\begin{subequations}
\be
\psi^{(1,2)}
(r,\tilde{\theta})=\varphi_0^0\left(r\left|\sin({\tilde{\theta}}/{2})\right|\right)
{\cal F}_\epsilon^{1,2}\left(r\cos({\tilde{\theta}}/{2})\right),
\label{asymptoticfunctiona}
\ee
where ``local'' coordinates near $\n_{1/2}$ on the unit sphere are introduced as 
$
\n(\theta,\phi)=\cos\tilde{\theta}\n_i
+ \sin\tilde{\theta}\cos\tilde{\phi}\n_i^\prime
+\sin\tilde{\theta}\sin\tilde{\phi}\n_i^{\prime\prime},$
where $\n_i^\prime$ and $\n_i^{\prime\prime}$ are two unit vectors orthogonal to each other
and to $\n_i$. Representation \rref{asymptoticfunctiona} is valid
 if the tunneling between the two minima is weak.
Substituting \req{asymptoticfunctiona} into \req{problemchannel}, 
treating the singular logarithmic potential exactly
and the remainder in second-order perturbation theory, we find  
\begin{align}
&\left[\frac{1}{x}\frac{d}{dx}x
\frac{d}{dx}+\frac{\gamma_{1,2}^2}{x^2}\right]{\cal F}_{\epsilon}^{(1,2)}(x)=\tilde{\epsilon}^{(1,2)} {\cal F}_{\epsilon}^{(1,2)}(x);
\label{asymptoticfunctionb}
\\
&-\tilde{\epsilon}^{(1,2)}\equiv \epsilon-\chi_0^0-
\frac{1}{2}\ln\frac{2\sin^2\frac{\theta_1+\theta_2}{2}}{\sin^2\frac{\theta_{1,2}}{2}},
\nonumber
\end{align}
where $\tilde{\epsilon}$ is the binding energy of a complex and dimensionless strength of the van der Waals attraction is 
\be
\begin{split}
& \gamma_{1,2}^2=1.23 \left[\cot({\theta_{1,2}}/{2})
-\cot(({\theta_1+\theta_2})/{2})\right]^2
\\
&\qquad \equiv 1.23 \mu_{2,1}(\mu_{1,2}+\mu_3)^2/[M\mu_{1,2}\mu_3].
\end{split}
\label{gammafinal}
\ee

\begin{figure}[h]
\includegraphics[width=0.6\columnwidth]{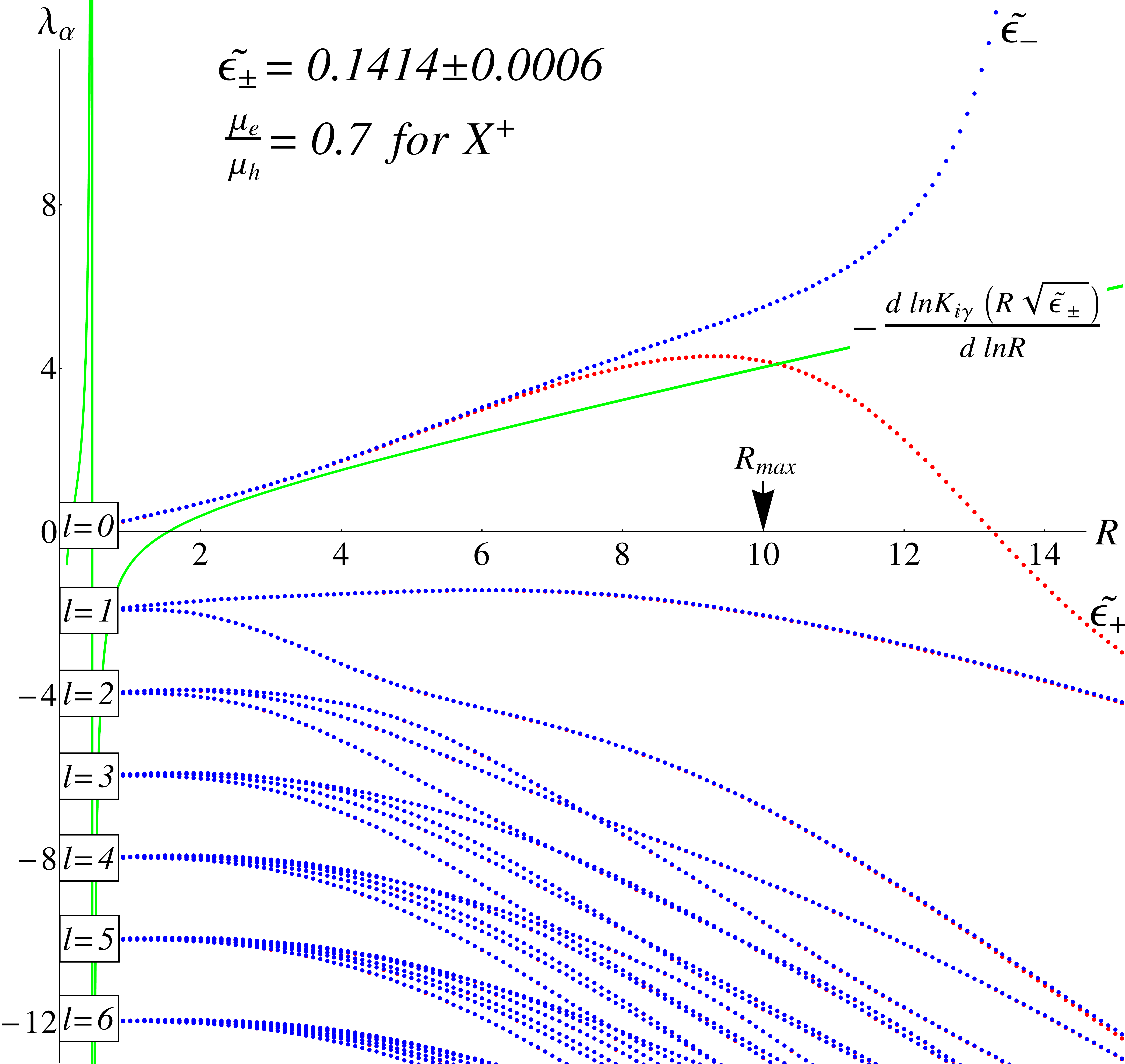}
\caption{Eigenvalues $\lambda_{\alpha}$ of matching matrix $\hLambda (R)$ evaluated numerically  
for $\epsilon=\epsilon_{\pm}$ slightly above (blue, $\epsilon_-$) and slightly below (red, $\epsilon_+$) 
the bound state energy of a trion with $\mu_e=0.7\mu_h$. 
The energy-sensitive highest eigenvalue is compared to the asymptotic of a logarithmic derivative in Eq. \rref{F} (green) 
calculated at the converged binding energy. }
\label{fig2}
\end{figure}

The solution corresponding to the bound state is
\be
{\cal F}^{(1,2)}=K_{i\gamma_{1,2}}\left(x\sqrt{\tilde{\epsilon}^{(1,2)}}\right),
\label{F}
\ee
where $K_{i\gamma}(y)$ is the MacDonald function, 
$\tilde{\epsilon}^{(1,2)}$ is determined by  matching 
\req{F} with the solution of \req{problemchannel}
\be
\psi_{l,m}(r)\propto r^{2l}, \quad r\ll 1.
\label{smallr}
\ee
In the interaction region, $ r\simeq 1$, the problem can only 
be handled numerically. Numerical
solution of \req{problemchannel} is not practical as many states
in the interaction region are evanescent (see Fig~\ref{fig1}),
and the search for the bound state would require the finding
of $N=(L_{max}+1)(L_{max}+2)/2$ boundary conditions at $r \to 0$ with
exponential accuracy.  Instead, we employ a procedure
that does not suffer from exponential dependence on $r$.
\label{asymptoticfunction}
\end{subequations}

We notice that one can replace solving \req{problemchannel} for
all $r$, with the solution on only $r>R$, where $R>0$ is an arbitrary
distance, if one knows the $N\times N$ {\em boundary condition matrix} $\hLambda$ 
defining the behavior of $\bpsi(r\to R+0)$, 
\be
\left[r{d\bpsi}/{dr}+ \hLambda(R) \bpsi \right]_{r=R}=0.
\label{Mconnector1}
\ee

\begin{subequations}
Requiring the invariance of solutions of 
\req{problemchannel} with respect to changes in $R$, we find
\be
R{d\hLambda}/{d R}=
R^2\left[{\epsilon\openone}-\hat{H}(R)\right]-2\hLambda+\hLambda^2,
\label{RGequationM}
\ee
where matrix multiplication is defined in the basis \rref{basis} as
$\left[\hat{A}\hat{B}\right]_{l_1m_1}^{l_2m_2}= 
\sum_{l=0}^{L_{max}}\sum_{m=0}^l\left[\hat{A}\right]_{l_1m_1}^{lm}
\left[\hat{B}\right]_{lm}^{l_2m_2}$, $\hLambda=\hLambda^*=\hLambda^\dagger$, and the initial condition follows from \req{smallr},
\be
\left[\hLambda(0)\right]_{lm}^{l'm'}=-2 l\delta_{ll'}\delta_{mm'}.
\label{RGinitialM}
\ee
\end{subequations}

The asymptotic dependence of the highest eigenvalues $\lambda_{\alpha}$ of matrix $\hLambda (R)$ 
corresponds to the asymptotic wave function in 
\reqs{asymptoticfunctiona} and \rref{F}, so that for an energy $\epsilon$ corresponding to a bound state \cite{FoootnoteC},
\be
\lambda_{\alpha}(R\gg 1)=\left.-
(x/{\cal F}) d {\cal F}/{dx}\right|_{x=R}.
\label{finalmatching}
\ee
We use \req{finalmatching} to find  energies of bound states numerically. 
First, we match tangentially the numerically calculated 
dependence of the highest eigenvalue $\lambda_{0}(R)$ using Eq. \rref{finalmatching} (as illustrated in Fig. 2), and find distance $R^{(i)}$
and an overestimated binding energy $\tilde{\epsilon}^{(i)}$. Next, we choose a distance $R_{max}$, $R^{(i)}<R_{max}<L_{max}$, 
to be used as a reference point in the rest of iterative procedure. Then, using Eqs. \rref{RGequationM} 
with $1/R_{\epsilon}^2$ determined by the variable energy $\tilde{\epsilon}<\tilde{\epsilon}^{(i)}$, we evaluate $\hLambda (R_{max})$, 
and its highest eigenvalue $\lambda_{\alpha}^0(R_{max})$, and find such energy $\tilde{\epsilon}$ 
that $\left[ \left. x d {\cal F}/({\cal F}{dx})\right|_{x=R_{max}}+\lambda_{\alpha}(R_{max})\right]^2$
is minimal. 
The outcome of such matching is examplified in Fig.~\ref{fig2} showing the  
eigenvalues of matrix $\hLambda (R)$ found for a trion \cite{FootnoteA}.

The resulting binding energies, calculated for various cases listed in Table I and various electron/hole mass ratios \cite{footnote} 
are shown Fig.~\ref{fig3} where, for comparison, we also show our results of the trion binding energies calculated using the diffision quantum Monte Carlo method. 
These two theoretical approaches give very close values, within the error bars determined by the limited size, $L_{max}$ of the spherical-harmonic basis. 
This agreement indicates that the new method offers an efficient tool to study complexes with more generic forms of electron-electron 
and electron-hole interaction, taking into account crossover from logarithmic to $1/r$ dependence at the longest distances. Note that the results displayed in Fig.~\ref{fig3} 
for $\mu_e<\mu_h$ can be used for $\mu_h<\mu_e$ by swapping $(\mu_e,X^+,X^A) \leftrightarrow (\mu_h,X^-,X^D)$.

After comparing the binding energies of various three-particle complexes, we conclude that the 'third' charge is more weakly bound 
(has a smaller dissociation energy) in an exciton localised on a charged donor or acceptor than in a trion \cite{FootnoteB}. 
As a result, heating of 2D crystal would suppress the luminescence from localised complexes much more than the luminescence of trions, because 
the evaporation one of the optically active carriers from $X^{D(A)}$ would happen at a much lower teperature than the temperature required for the decomposition of $X^{\pm}$. 
Such behavior is highly counter-intuitive, because, despite weaker binding, the line of $X^{D/A}$ in recombination spectra,
$\omega_{X^{D(A)}}=\omega_X-\frac{e^2}{r_*}\left(\tilde{\epsilon}_{X^{D(A)}}+\frac12\ln\left[1+\frac{\mu_{e(h)}}{\mu_{h(e)}}\right]\right)$, lies below (red-shifted) the line of a trion, 
 $\omega_{X^{\pm}}=\omega_X-\frac{e^2}{r_*}\tilde{\epsilon}_{X^{\pm}}$. For comparable masses of electrons and holes, 
the exciton-trion splitting appears to be an order of magnitude smaller than the splitting between the ground state of the exciton and its first optically active excited state $X_1^0$, 
at $\Delta_1=\omega_{X^*}-\omega_X=1.14\frac{e^2}{r_*}$, whereas $\omega_{X^{D(A)}}-\omega_X\simeq 0.5\Delta_1 $, as prescribed by the 
the two-particle binding energy of electon/hole in donor/acceptor being much larger than the one of the exciton, 
overcompensating the difference between the three-particle binding energies. Such temperature behavior of the lower end of recombination spectra in TMDCs has recently been 
observed in several experiments on WSe$_2$ \cite{Zhu,HeinzU,PotemskiU}.

We thank V. Cheianov, T. Heinz, A. Morpurgo  for discussions, and D. Gradinar for help in preparing the manuscript. This work was supported
by EC FP7 {\it Graphene Flagship} project CNECT-ICT-604391, ERC Synergy {\it Grant Hetero2D}, and the Simons Foundation.

\ifbool{arxive}{

\clearpage

\newpage*

\includepdf[pages=1]{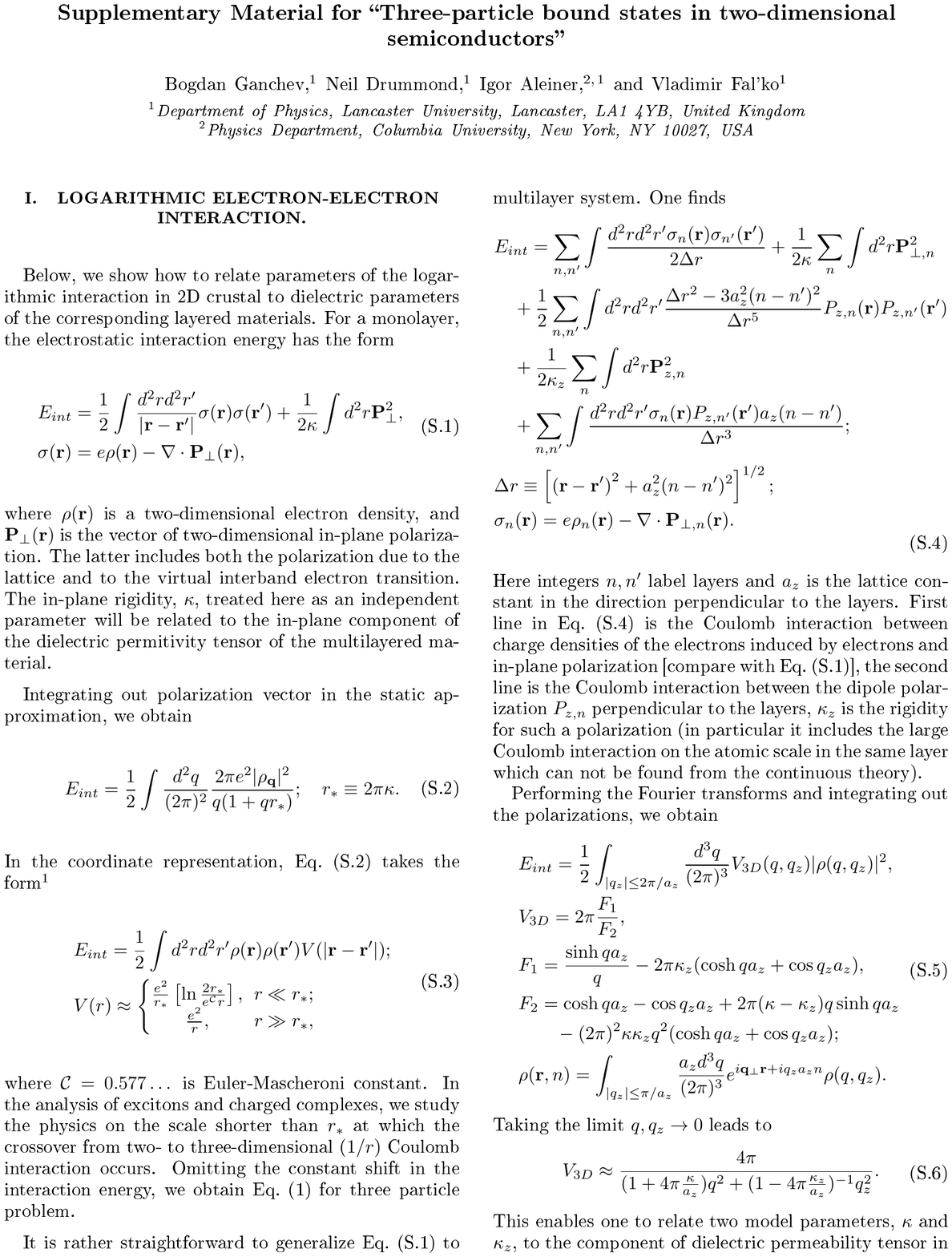}

\newpage*

\includepdf[pages=2]{SupplementalMaterial.pdf}
\newpage*

\includepdf[pages=3]{SupplementalMaterial.pdf}

\newpage*

\includepdf[pages=4]{SupplementalMaterial.pdf}

\newpage*



\section*{Supplementary material}

\* **Supplementary Material for "Three-particle bound states in two-dimensional semiconductors"**

Bogdan Ganchev,[1] Neil Drummond,[1] Igor Aleiner,[2,1] and Vladimir Fal'ko[1]

[1]*Department of Physics, Lancaster University, Lancaster, LA1 4YB, United Kingdom*
[2]*Physics Department, Columbia University, New York, NY 10027, USA*

## I. LOGARITHMIC ELECTRON-ELECTRON INTERACTION.

Below, we show how to relate parameters of the logarithmic interaction in 2D crustal to dielectric parameters of the corresponding layered materials. For a monolayer, the electrostatic interaction energy has the form

$$E_{int} = \frac{1}{2}\int \frac{d^2r d^2r'}{|\mathbf{r}-\mathbf{r}'|}\sigma(\mathbf{r})\sigma(\mathbf{r}') + \frac{1}{2\kappa}\int d^2r \mathbf{P}_\perp^2, \quad (S.1)$$
$$\sigma(\mathbf{r}) = e\rho(\mathbf{r}) - \nabla\cdot\mathbf{P}_\perp(\mathbf{r}),$$

where $\rho(\mathbf{r})$ is a two-dimensional electron density, and $\mathbf{P}_\perp(\mathbf{r})$ is the vector of two-dimensional in-plane polarization. The latter includes both the polarization due to the lattice and to the virtual interband electron transition. The in-plane rigidity, $\kappa$, treated here as an independent parameter will be related to the in-plane component of the dielectric permitivity tensor of the multilayered material.

Integrating out polarization vector in the static approximation, we obtain

$$E_{int} = \frac{1}{2}\int \frac{d^2q}{(2\pi)^2}\frac{2\pi e^2 |\rho_\mathbf{q}|^2}{q(1+qr_*)}; \quad r_* \equiv 2\pi\kappa. \quad (S.2)$$

In the coordinate representation, Eq. (S.2) takes the form[1]

$$E_{int} = \frac{1}{2}\int d^2r d^2r' \rho(\mathbf{r})\rho(\mathbf{r}')V(|\mathbf{r}-\mathbf{r}'|);$$
$$V(r) \approx \begin{cases} \frac{e^2}{r_*}\left[\ln\frac{2r_*}{e^\mathcal{C} r}\right], & r \ll r_*; \\ \frac{e^2}{r}, & r \gg r_*, \end{cases} \quad (S.3)$$

where $\mathcal{C} = 0.577\ldots$ is Euler-Mascheroni constant. In the analysis of excitons and charged complexes, we study the physics on the scale shorter than $r_*$ at which the crossover from two- to three-dimensional ($1/r$) Coulomb interaction occurs. Omitting the constant shift in the interaction energy, we obtain Eq. (1) for three particle problem.

It is rather straightforward to generalize Eq. (S.1) to multilayer system. One finds

$$E_{int} = \sum_{n,n'}\int \frac{d^2r d^2r' \sigma_n(\mathbf{r})\sigma_{n'}(\mathbf{r}')}{2\Delta r} + \frac{1}{2\kappa}\sum_n\int d^2r \mathbf{P}_{\perp,n}^2$$
$$+ \frac{1}{2}\sum_{n,n'}\int d^2r d^2r' \frac{\Delta r^2 - 3a_z^2(n-n')^2}{\Delta r^5}P_{z,n}(\mathbf{r})P_{z,n'}(\mathbf{r}')$$
$$+ \frac{1}{2\kappa_z}\sum_n\int d^2r \mathbf{P}_{z,n}^2$$
$$+ \sum_{n,n'}\int \frac{d^2r d^2r' \sigma_n(\mathbf{r})P_{z,n'}(\mathbf{r}')a_z(n-n')}{\Delta r^3};$$
$$\Delta r \equiv \left[(\mathbf{r}-\mathbf{r}')^2 + a_z^2(n-n')^2\right]^{1/2};$$
$$\sigma_n(\mathbf{r}) = e\rho_n(\mathbf{r}) - \nabla\cdot\mathbf{P}_{\perp,n}(\mathbf{r}). \quad (S.4)$$

Here integers $n, n'$ label layers and $a_z$ is the lattice constant in the direction perpendicular to the layers. First line in Eq. (S.4) is the Coulomb interaction between charge densities of the electrons induced by electrons and in-plane polarization [compare with Eq. (S.1)], the second line is the Coulomb interaction between the dipole polarization $P_{z,n}$ perpendicular to the layers, $\kappa_z$ is the rigidity for such a polarization (in particular it includes the large Coulomb interaction on the atomic scale in the same layer which can not be found from the continuous theory).

Performing the Fourier transforms and integrating out the polarizations, we obtain

$$E_{int} = \frac{1}{2}\int_{|q_z|\leq 2\pi/a_z}\frac{d^3q}{(2\pi)^3}V_{3D}(q,q_z)|\rho(q,q_z)|^2,$$
$$V_{3D} = 2\pi\frac{F_1}{F_2},$$
$$F_1 = \frac{\sinh qa_z}{q} - 2\pi\kappa_z(\cosh qa_z + \cos q_z a_z), \quad (S.5)$$
$$F_2 = \cosh qa_z - \cos q_z a_z + 2\pi(\kappa-\kappa_z)q\sinh qa_z$$
$$- (2\pi)^2\kappa\kappa_z q^2(\cosh qa_z + \cos q_z a_z);$$
$$\rho(\mathbf{r},n) = \int_{|q_z|\leq \pi/a_z}\frac{a_z d^3q}{(2\pi)^3}e^{i\mathbf{q}_\perp\cdot\mathbf{r}+iq_z a_z n}\rho(q,q_z).$$

Taking the limit $q, q_z \to 0$ leads to

$$V_{3D} \approx \frac{4\pi}{(1+4\pi\frac{\kappa}{a_z})q^2 + (1-4\pi\frac{\kappa_z}{a_z})^{-1}q_z^2}. \quad (S.6)$$

This enables one to relate two model parameters, $\kappa$ and $\kappa_z$, to the component of dielectric permeability tensor in

the bulk of the layered material,

$$\varepsilon_\parallel = 1 + 4\pi\frac{\kappa}{a_z}, \qquad \varepsilon_z = \frac{1}{1 - 4\pi\frac{\kappa_z}{a_z}}, \qquad (S.7)$$

This relation enables us to represent $r_* = a_z(\varepsilon_\parallel - 1)/2$. Note that $\kappa_z$ includes the short range Coulomb interaction within the same layer, so that the inequality $4\pi\kappa_z < a_z$ always holds.

To describe the interaction between two charges at the intermediate distances within the same 2D layer of the bulk crystal, one need to use an effective 2D interaction,

$$V = \int_{-\pi/a}^{\pi/a} \frac{dq_z}{2\pi} V_{3D}(qa \ll 1, q_z) \approx \frac{e^2}{a\kappa q^2} = \frac{4\pi}{(\epsilon_\parallel - 1)q^2}, \qquad (S.8)$$

which results in that the logarithmic approximation, (S.3) is also applicable to the description of the Coulomb interaction in the bulk of layered material

$$V(r) \approx \frac{e^2}{r_*} \ln \frac{r_*}{r\sqrt{\varepsilon_z\varepsilon_\parallel}}; \qquad r < r_*/\sqrt{\varepsilon_z\varepsilon_\parallel}. \qquad (S.9)$$

At larger $r$ it matches the asymptotic tail $e^2/r\sqrt{\varepsilon_z\varepsilon_\parallel}$ corresponding Eq. (S.6).

## II. COORDINATES TRANSFORMATION AND SEPARATION OF VARIABLES.

To perform the coordinate transformation leading to Eq. (3) in the main text, we rewrite the four-dimensional Laplacian in the covariant form

$$\nabla_4^2 \equiv \nabla_{\tilde{\mathbf{r}}}^2 + \nabla_{\mathbf{r}'}^2 = \frac{1}{\sqrt{g}} \nabla_i \sqrt{g} g^{ij} \nabla_j \quad g = \det \hat{g}, \quad (S.10)$$

where $g_{ij}$ is the metric tensor $g_{ik}g^{kj} = \delta_i^j$, and in the original coordinates $\hat{g} = \mathbb{1}$.

Coordinates introduced in the main text

$$\begin{pmatrix} r'_x \\ r'_y \\ \tilde{r}_x \\ \tilde{r}_y \end{pmatrix} = r \begin{pmatrix} \cos\frac{\theta}{2}\cos\left(\Phi + \frac{\phi}{2}\right) \\ \cos\frac{\theta}{2}\sin\left(\Phi + \frac{\phi}{2}\right) \\ \sin\frac{\theta}{2}\cos\left(\Phi - \frac{\phi}{2}\right) \\ \sin\frac{\theta}{2}\sin\left(\Phi - \frac{\phi}{2}\right) \end{pmatrix}, \qquad (S.11)$$

correspond to the metric tensor

$$\hat{g} = \begin{pmatrix} 1 & 0 & 0 & 0 \\ 0 & \frac{r^2}{4} & 0 & 0 \\ 0 & 0 & \frac{r^2}{4} & \frac{r^2}{2}\cos\theta \\ 0 & 0 & \frac{r^2}{2}\cos\theta & r^2 \end{pmatrix}. \qquad (S.12)$$

Substitution of Eq. (S.12) into Eq. (S.10) gives Eq. (3) of the main text.

The use of this metric tensor results in the form of the Schrödinger equation in Eq. (3) in the main text,

$$\begin{aligned} &\left[-\nabla_4^2 + \ln r + U(\theta, \phi)\right]\psi = \epsilon\psi; \\ &-\nabla_4^2 = -\frac{\partial^2}{\partial r^2} - \frac{3\partial}{r\partial r} + \frac{4\hat{\mathbf{L}}^2}{r^2} + \frac{\hat{\Theta}}{r^2\sin^2\theta}; \\ &U(\theta,\phi) = \frac{1}{2}\ln\left[\frac{(1 - \mathbf{n} \cdot \mathbf{n}_1)(1 - \mathbf{n} \cdot \mathbf{n}_2)}{(1 - \mathbf{n} \cdot \mathbf{n}_z)}\right], \\ &\mathbf{n} = [\sin\theta\cos\phi; \ \sin\theta\sin\phi; \ \cos\theta], \end{aligned} \qquad (S.13)$$

with

$$\hat{\mathbf{L}}^2 = -\frac{1}{\sin\theta}\frac{\partial}{\partial\theta}\sin\theta\frac{\partial}{\partial\theta} - \frac{1}{\sin^2\theta}\frac{\partial^2}{\partial\phi^2}$$

and

$$\hat{\Theta} = \left[-\frac{\partial}{\partial\Phi} + 4\cos\theta\frac{\partial}{\partial\phi}\right]\frac{\partial}{\partial\Phi}.$$

Because of the rotational symmetry, potential $U$ in Eq. (S.13) does not depend on the angle $\Phi$. Therefore, the eigenstates can be classified by the integer angular momentum $J$ ( $\pm J$ are degenerate due to the time inversion):

$$\Psi_J(r, \theta, \phi, \Phi) = e^{iJ\Phi}\psi_J(r, \theta, \phi). \qquad (S.14a)$$

For the wavefunction $\Psi$ to be single-valued,

$$\begin{aligned} &\psi_J(r, \theta + 2\pi, \phi) = \psi_J(r, \theta, \phi + 2\pi) = (-1)^J\psi_J(r, \theta, \phi); \\ &\psi_J(r, -\theta, \phi + \pi) = (i)^J\psi_J(r, \theta, \phi). \end{aligned} \qquad (S.14b)$$

Potential $U$ has a mirror reflection symmetry $U(\phi) = U(-\phi)$ (hereinafter we will omit coordinates invariant under transformations), which distinguishes the states into the groups of even ($e$) and odd ($o$) states

$$\psi_J^e(\phi) = \psi_{-J}^e(-\phi); \ \psi_J^o(\phi) = -\psi_{-J}^o(-\phi). \qquad (S.14c)$$

If two particles in the complex are identical, $\theta_1 = \theta_2$,

$$U(\theta) = U(-\theta), \qquad (S.14d)$$

the states split into symmetric/anti-symmetric ($s/a$):

$$\psi_J^s(-\theta) = \psi_J^s(\theta); \quad \psi_J^a(-\theta) = -\psi_J^a(\theta). \qquad (S.14e)$$

This symmetry determines spin/valley multiplets discussed in Section IV.

The resulting reduction of the number of relevant degreees of freedom needed to describe ground states of three-particle complexes enables us to employ a computationally inexpensive method alternative to the Faddeev equation[4] and variational functions approaches[5,6] commonly used in atomic physics.

## III. MATRIX ELEMENTS OF THE INTERACTION AND THE STRUCTURE OF ANGULAR HAMILTONIAN.

The derivation of matrix elements of 'dimensionless potential' $U(\mathbf{n})$ consists in the evaluation of the integral

$$\tilde{U}^{l_1 m_1}_{l_2 m_2} \equiv \int_0^\pi d\theta \sin\theta \int_0^{2\pi} d\phi \, U(\mathbf{n}) Y^*_{l_1 m_1}(\mathbf{n}) Y_{l_2 m_2}(\mathbf{n}), \quad (S.15)$$

where the spherical harmonics $Y_{lm}(\mathbf{n})$ are defined according to convention of Ref. 3, and the potential $U$ is defined in Eq. (3) of the main text. Using the formula[2],

$$\ln(1 - \mathbf{n} \cdot \mathbf{n}_1) = -\sum_{l=1}^\infty \frac{2l+1}{l(l+1)} P_l(\mathbf{n}\cdot\mathbf{n}_1) + \ln 2 - 1, \quad (S.16)$$

applying addition theorem for the spherical harmonics

$$P_l(\mathbf{n} \cdot \mathbf{n}_1) = \frac{4\pi}{2l+1} \sum_{m=-l}^{l} Y^*_{l,m}(\mathbf{n}_1) Y_{l,m}(\mathbf{n}),$$

and the integral relation leading to $3j$-symbols[3],

$$\int_0^\pi d\theta \sin\theta \int_0^{2\pi} d\phi \, Y_{l_1 m_1}(\mathbf{n}) Y_{l_2 m_2}(\mathbf{n}) Y_{l_3 m_3}(\mathbf{n})$$
$$= \left[\frac{\prod_{i=1,2,3}(2l_i+1)}{4\pi}\right]^{1/2} \begin{pmatrix} l_1 & l_2 & l_3 \\ m_1 & m_2 & m_3 \end{pmatrix} \begin{pmatrix} l_1 & l_2 & l_3 \\ 0 & 0 & 0 \end{pmatrix}, \quad (S.17)$$

where $m_1 + m_2 + m_3 = 0$ and $l_{1,2,3}$ obey the triangular inequality, we arrive at

$$\tilde{U}^{l_1 m_1}_{l_2 m_2} = \frac{1}{2}(\ln 2 - 1)\delta_{l_1 l_2}\delta_{m_1 m_2} - [\pi(2l_1+1)(2l_2+1)]^{1/2}(-1)^{m_1}$$
$$\times \sum_{l=\max(1,|l_1-l_2|,|m_1-m_2|)}^{l_1+l_2} \frac{\sqrt{2l+1}}{l(l+1)} \begin{pmatrix} l_1 & l & l_2 \\ -m_1 & m_1-m_2 & m_2 \end{pmatrix} \begin{pmatrix} l_1 & l & l_2 \\ 0 & 0 & 0 \end{pmatrix} \left[Y^*_{l,m_1-m_2}(\mathbf{n}_1) + Y^*_{l,m_1-m_2}(\mathbf{n}_2) - Y^*_{l,m_1-m_2}(\mathbf{n}_z)\right]. \quad (S.18)$$

Here, unit vector $\mathbf{n}_{1,2,z}$ are defined in Eq. (4) of the main text. Symmetrizing matrix element according to Eq. (6) for $Y^e$, we obtain Eq. (8) and the following non-numbered equation of the main text.

## IV. OPTICALLY ACTIVE TRION STATES

In monolayers of transition metal dichalcogenides the absolute minima of conduction and and absolute maxima of the valence bands are at the two non-equivalent Brillouin zone corners $K, K'$, known as valleys. The atomic spin-orbit interaction in transition metals introduces large spin-orbit splitting of the valence band $(v)$[7–9,11] so that the low-energy excitons and trions involve only the lowest hole states,

$$|+,h\rangle = |K,\uparrow,v\rangle; \quad |-,h\rangle = |K',\downarrow,v\rangle. \quad (S.19)$$

In contrast, the spin orbit interaction for electrons is much smaller than the trion binding energy so that for each electron 4 states coming from the spin and valley degeneracy have to be taken into account.

### A. Negatively charged trion, $X^-$

As we discussed in the main text, only symmetric orbital state is relevant (other bound states have exponentially small binding energy), so that we are left only with the six degenerate states anti-symmetric with respect to the permutations of the valley and spin indices (below, $c$ denotes the conduction band):

$$\sqrt{2}\,|+,ee\rangle = |K',\downarrow,c\rangle\,|K,\uparrow,c\rangle - |K,\uparrow,c\rangle\,|K',\downarrow,c\rangle;$$
$$\sqrt{2}\,|-,ee\rangle = |K,\downarrow,c\rangle\,|K',\uparrow,c\rangle - |K',\uparrow,c\rangle\,|K,\downarrow,c\rangle;$$
$$\sqrt{2}\,|1,ee\rangle = |K,\downarrow,c\rangle\,|K,\uparrow,c\rangle - |K,\uparrow,c\rangle\,|K,\downarrow,c\rangle;$$
$$\sqrt{2}\,|2,ee\rangle = |K',\downarrow,c\rangle\,|K',\uparrow,c\rangle - |K',\uparrow,c\rangle\,|K',\downarrow,c\rangle;$$
$$\sqrt{2}\,|3,ee\rangle = |K,\uparrow,c\rangle\,|K',\uparrow,c\rangle - |K',\uparrow,c\rangle\,|K,\uparrow,c\rangle;$$
$$\sqrt{2}\,|4,ee\rangle = |K,\downarrow,c\rangle\,|K',\downarrow,c\rangle - |K',\downarrow,c\rangle\,|K,\downarrow,c\rangle. \quad (S.20)$$

Spin-orbit interaction splits single electron states into two Kramers doublets: $E_{K,\uparrow} = E_{K',\downarrow} = \epsilon_{so}$, $E_{K',\uparrow} = E_{K,\downarrow} = -\epsilon_{so}$. Accordingly, sextuplet of two electron states (S.20) is split into two singlets $E_{\pm,ee} = \pm 2\epsilon_{so}$ and a quadruplet $E_{1-4,ee} = 0$. The quadruplet may be further split due to the lattice effects (i.e. trigonal warping) which we neglect. Also, we neglect the electron-hole exchange. Then the spin-valley trion states are direct product of the two-electron (S.20) and hole states (S.19).

Because the optical transition conserves the spin and the quasi-momentum, the allowed optical transition by the circular left-handed polarized light are

$$|+,ee\rangle\,|+,h\rangle \to |K,\uparrow,c\rangle;$$
$$|2,ee\rangle\,|+,h\rangle \to |K',\uparrow,c\rangle; \quad (S.21)$$
$$|4,ee\rangle\,|+,h\rangle \to |K,\downarrow,c\rangle;$$

and

$$|-, ee\rangle |-, h\rangle \to |K', \downarrow, c\rangle ;$$
$$|1, ee\rangle |-, h\rangle \to |K, \downarrow, c\rangle ; \quad \text{(S.22)}$$
$$|3, ee\rangle |-, h\rangle \to |K', \uparrow, c\rangle ;$$

for the right-handed polarization. It is important to emphasize that even though the trion states and single-electron states are split by spin-orbit coupling, the optical line is not split, similarly to the exciton.

### B. Positively charged trion, $X^+$.

As in the previous subsection, we are interested in the orbital symmetric state. The only allowed spin/valley part of the two-hole system is (see Eq. (S.19))

$$\sqrt{2}\, |hh\rangle = |K', \downarrow, v\rangle |K, \uparrow, v\rangle - |K, \uparrow, v\rangle |K', \downarrow, v\rangle . \quad \text{(S.23)}$$

The allowed optical transition by the circular left-handed polarized light are

$$|hh\rangle |K, \uparrow, c\rangle \to |K, \uparrow, v\rangle , \quad \text{(S.24)}$$

and

$$|hh\rangle |K', \downarrow, c\rangle \to |K', \downarrow, v\rangle , \quad \text{(S.25)}$$

for the right-handed polarization. Once again, these lines are not split by spin-orbit interaction.

## V. DIFFUSION QUANTUM MONTE CARLO CALCULATIONS.

To confirm the accuracy of the results obtained using the novel method described above, we have carried out diffusion quantum Monte Carlo (DMC) calculations[12,13] of the exciton and trion ground-state energies for a range of electron–hole mass ratios. Similar calculations have previously been performed for 2D biexcitons interacting via a $1/r$ potential[14,15]. The ground-state wave functions of the exciton and trion are nodeless and hence the fixed-node DMC energy is exact, which gives it a preference over the Faddeev equation[4] and variational functions approaches[5,6] commonly used in atomic physics.

We have performed numerical calculations of the ground-state energies of excitons and trions using the variational and diffusion quantum Monte Carlo (VMC and DMC) methods[12,16]. The logarithmic interaction shown in Eq. (1) of main text was used in our calculations. Our trial wave functions were of form $\Psi = \exp(J)$, where the Jastrow exponent $J$ contained pairwise terms

$$u_{\rm eh}(r) = c_1 r^2 \log(r) \exp(-c_2 r^2) - [1 - \exp(-c_2 r^2)]c_3 r \quad \text{(S.26)}$$

between electrons and holes and

$$u_{\rm ee}(r) = c_4 r^2 \log(r) \exp(-c_5 r^2) \quad \text{(S.27)}$$

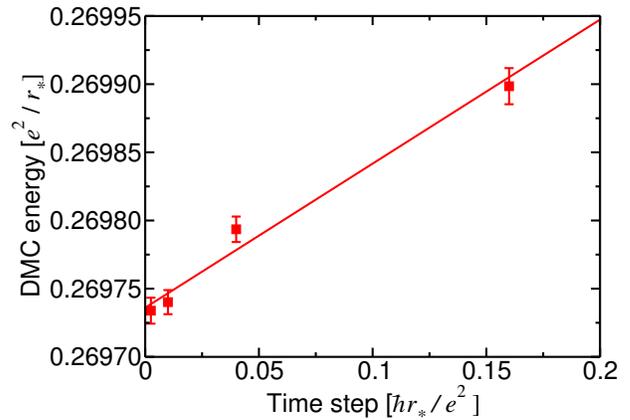

Figure 1. Extrapolation of the DMC total energy of a trion (equal-mass case) to zero time step and infinite population.

between electrons, where $c_1$, $c_2$, $c_3$, $c_4$, and $c_5$ are parameters. In order to satisfy the analog of the Kato cusp conditions (i.e., to make the local energy nondivergent at coalescence points), we must have $c_1 = e^2 \mu_{\rm e} \mu_{\rm h}/[2\hbar^2 r_*(\mu_{\rm e} + \mu_{\rm h})]$ and $c_4 = -e^2 \mu_{\rm e}/(4\hbar^2 r_*)$. For $\Psi$ to be normalizable we must also have $c_2 > 0$, $c_3 > 0$, and $c_5 > 0$. In addition, the Jastrow exponent $J$ contained cuspless polynomials in the interparticle distances, which were truncated smoothly at finite range, together with a three-body polynomial term coupling the interparticle distances in the trion[17,18]. The free parameters in our trial wave function were optimized by unreweighted variance minimization[19,20] and energy minimization[21].

Our DMC calculations were performed using time steps in the ratio 1:4, with the corresponding target configuration populations being in the ratio 4:1. The resulting DMC energies were extrapolated linearly to zero time step and infinite population. An example of the DMC time-step bias is shown in Fig. 1. It is clear that the bias is linear in the time step, as expected. The bias is small in any case: see the energy scale in Fig. 1.

The pair-distribution function (PDF) between particles $i$ and $j$ in an exciton or trion is defined as

$$g_{ij}(r) = \frac{1}{2\pi r} \langle \delta(r - |\hat{\mathbf{r}}_i - \hat{\mathbf{r}}_j|) \rangle . \quad \text{(S.28)}$$

In a negative trion, the electron–hole PDF is defined as the sum of the electron–hole PDFs for spin-up and spin-down electrons. We evaluated the extrapolated estimate of the PDF (twice the DMC mixed estimate minus the VMC estimate). The error in the extrapolated estimate is second order in the error in the trial wave function[16].

PDF results are shown in Fig. 2. The relatively long range of the extent of the trion wave function, as compared to the exciton, is clear.

4



}
{}

\begin{thebibliography}{99}

\bibitem{TMDC1} 
B. Radisavljevic, A. Radenovic, J. Brivio, V. Giacometti, and A. Kis, Nature Nanotech. {\bf 6}, 147 (2011).  
\bibitem{TMDC2}
D. Lembke and A. Kis, ACS Nano {\bf 6}, 10070 (2012). 
\bibitem{TMDC3}
H. Wang, L. Yu, Y.H. Lee, Y. Shi, A. Hsu, M. Chin, L.J. Li, M. Dubey, J. Kong, and T. Palacios, Nano Letters {\bf 12}, 4674 (2012).
\bibitem{TMDC4}
X. Xu,	W. Yao, D. Xiao, and T.F. Heinz, Nature Phys. {\bf 10}, 343 (2014).

\bibitem{2Doptical1}
D. Xiao, G.B. Liu, W. Feng, X. Xu, and W. Yao, Phys. Rev. Lett. {\bf 108}, 196802 (2012).
\bibitem{2Doptical2}
H. Zeng, J. Dai, W. Yao, D. Xiao, and X. Cui, Nature Nanotech. {\bf 7}, 490 (2012).
\bibitem{2Doptical3}
K.F. Mak, K. He, J. Shan, and T.F. Heinz, Nature Nanotech. {\bf 7}, 494 (2012).
\bibitem{2Doptical4}
G. Sallen, L. Bouet, X. Marie, G. Wang, C.R. Zhu, W.P. Han, Y. Lu, P.H. Tan, T. Amand, B.L. Liu, and B. Urbaszek, Phys. Rev. B {\bf 86}, 081301 (2012).
\bibitem{2Doptical5}
T. Cao, G. Wang, W. Han, H. Ye, C. Zhu, J. Shi, Q. Niu, P. Tan, E. Wang, B. Liu, and J. Feng, Nature Comm. {\bf 3}, 887 (2012).
\bibitem{2Doptical6}
H. Conley, B. Wang, J. Ziegler, R. Haglund, S. Pantelides, K.I. Bolotin, Nano Letters {\bf 13}, 3626 (2013).

\bibitem{2DBands1}
A. Kuc, N. Zibouche, and T. Heine, Phys. Rev. B {\bf 83}, 245213 (2011).
\bibitem{2DBands2}
Z. Y. Zhu, Y. C. Cheng, and U. Schwingenschl\"ogl, Phys. Rev. B {\bf 84}, 153402 (2011). 
\bibitem{2DBands3}
A. Molina-Sanchez, D. Sangalli, K. Hummer, A. Marini, and L. Wirtz,  Phys. Rev. B {\bf 88}, 045412 (2013). 
\bibitem{2DBands4}
W. Jin \emph{et al.}, Phys. Rev. Lett. {\bf 111}, 106801 (2013).
\bibitem{2DBands5}
A. Kormanyos, V. Zolyomi, N.D. Drummond, P. Rakyta, G. Burkard, and V.I. Falko,  Phys. Rev. B {\bf 88}, 045416 (2013).


\bibitem{Excitons1} 
A. Ramasubramaniam, Phys. Rev. B {\bf 86}, 115409 (2012).
\bibitem{Excitons2}
H.P. Komsa and A.V. Krasheninnikov, Phys. Rev. B {\bf 86}, 241201 (2012). 
\bibitem{Excitons3}
D.Y. Qiu, F.H. da Jornada, and S.G. Louie, Phys. Rev. Lett. {\bf 111}, 216805 (2013).
\bibitem{Excitons4}
M.M. Glazov,  T. Amand, X. Marie, D. Lagarde, L. Bouet, and B. Urbaszek, Phys. Rev. B 89, 201302 (2014).
\bibitem{Excitons5}
G. Berghauser and E. Malic, Phys. Rev. B {\bf 89}, 125309 (2014).
\bibitem{Excitons6}
A.R. Klots, A.K.M. Newaz, B. Wang, D. Prasai, H. Krzyzanowska, D. Caudel, N. J. Ghimire, J. Yan, B.L. Ivanov, K.A. Velizhanin, A. Burger, D.G. Mandrus, N.H. Tolk, S.T. Pantelides, and K.I. Bolotin, arXiv:1403.6455.

\bibitem{Trions1}
K.F. Mak, K. He, C. Lee, G.H. Lee, J. Hone, T.F. Heinz, J. Shan, Nature Mat. {\bf 12}, 207 (2013).
\bibitem{Trions2}
C. Zhang, H. Wang, W. Chan, C. Manolatou, and F. Rana, Phys. Rev. B {\bf 89}, 205436 (2014). 
\bibitem{Trions3}
A. Srivastava, M. Sidler, A.V. Allain, D.S. Lembke, A. Kis, and A. Imamoglu, arXiv:1407.2624.

\bibitem{footnote}
As shown in Ref. \cite{Sup}, $r_*=a_z(\varepsilon_{\parallel} -1)/2$, where $\varepsilon_{\parallel}$ is the in-plane component of the dielectric 
permitivity tensor of the bulk layered material and $a_z$ is the distance between layers in it. For example, 
in WS$_2$, this leads \cite{Heinz} to the estimate $r_*\sim 7.5$ nm, hence
$e^2/r_* \sim 200$ meV and a typical trion binding energy $\sim 30$ meV.


\bibitem{Heinz} 
A. Chernikov, T.C. Berkelbach, H.M. Hill, A. Rigosi, Y. Li, O.B. Aslan, D.R. Reichman, M.S. Hybertsen, and T.F. Heinz, Phys. Rev. Lett. {\bf 113}, 076802 (2014). 
\bibitem{Excitonlog1}
I.R. Lapidus, Am. J. Phys. {\bf 49}, 807 (1981).
\bibitem{Excitonlog2}
K. Eveker, D. Grow, B. Jost, C.E. Monfort, and K.W. Nelson, Am. J. Phys. {\bf 58}, 1183 (1990).

\bibitem{Ceperley_1980} D.~Ceperley and B.~Alder,   Phys. Rev. Lett. {\bf 45}, 566 (1980).
\bibitem{Needs_2010} R.~Needs {\it et al.}, J.~Phys.~Cond.~Matter  {\bf 22}, 023201 (2010).

\bibitem{2DBands7}
A. Kormanyos, G. Burkard, M. Gmitra, J. Fabian, V. Zolyomi, N. Drummond, V. Fal'ko, arXiv:1410.6666. 

\bibitem{Sup} {\em Supplementary Information}, which includes Refs. \cite{Keldysh,RG,Faddeev,variational1,variational2,Tan_2005,Lee_2009,Foulkes_2001,Drummond_2004,LopezRios_2012,Umrigar_1988,Drummond_2005,Umrigar_2007}

\bibitem{Keldysh}  L. V. Keldysh,  Pis'ma Zh. Eksp. Teor. Fiz. 30, 245 (1979)  (JETP Lett. 30, 224 (1979)).

\bibitem{RG}
L.S. Gradstein and I.M Ryzhik, {\em Tables of Integrals, series and Products}, 6th edition, Academic Press (2000).

\bibitem{Faddeev} 
L.D. Faddeev and S.P. Merkuriev, {\em Quantum Scattering Theory for Several Particle Systems}, Springer (1993).

\bibitem{variational1} 
E.A. Hylleraas, Z. Phys. {\bf 54}, 347 (1929).

\bibitem{variational2} 
S. Chandrasekhar, Astrophyical J. {\bf 100}, 176 (1944).

\bibitem{Tan_2005} 
M.Y.J. Tan, N.D. Drummond, and R.J. Needs, Phys. Rev. B \textbf{71}, 033303 (2005).

\bibitem{Lee_2009} 
R.M. Lee, N.D. Drummond, and R.J. Needs, Phys. Rev. B \textbf{79}, 125308 (2009).

\bibitem{Foulkes_2001} 
W.M.C. Foulkes, \textit{et al.}, Rev. Mod. Phys. \textbf{73}, 33 (2001).

\bibitem{Drummond_2004} 
N.D. Drummond, M.D. Towler, and R.J. Needs, Phys. Rev. B \textbf{70}, 235119 (2004).

\bibitem{LopezRios_2012} 
P. L\'{o}pez R\'{\i}os, \textit{et al.}, Phys. Rev. E \textbf{86}, 036703 (2012).

\bibitem{Umrigar_1988} 
C.J. Umrigar, K.G. Wilson, and J.W. Wilkins, Phys. Rev. Lett. \textbf{60}, 1719 (1988).

\bibitem{Drummond_2005} 
N.D. Drummond and R.J. Needs, Phys. Rev. B \textbf{72}, 085124 (2005).

\bibitem{Umrigar_2007} 
C.J. Umrigar, \textit{et al.},  Phys. Rev. Lett. \textbf{98}, 110201 (2007).

\bibitem{masses1} 
H. Shi, H. Pan, Y.-W. Zhang, and B.I. Yakobson, Phys. Rev. B {\bf 87}, 155304 (2013).
\bibitem{masses2} 
D. Wickramaratne, R.K. Lake, and F. Zahid, J. Chem. Phys. {\bf 140}, 124710 (2014).
\bibitem{masses3} 
N. Zibouche, P. Philipsen, T. Heine, and A. Kuc, arXiv:1403.0552.
\bibitem{masses4} 
W. Zhang, Z. Huang, and W. Zhang, arXiv:1403.3872. 

\bibitem{FootnoteParabolic}
Here, we neglect the inter-band mixing present in the {\bf k}$\cdotp${\bf p} theory Hamiltonians developed for the description 
of transition metal dichalcogenides \cite{2DBands5,2DBands7} which is important for the exciton-photon coupling 
and Berry curvature effects in transport \cite{QT1,QT2}, but is irrelevant for the analysis of exciton and trion binding energies 
since those are substantially less than the band gap. 

\bibitem{QT1}
M. Cazalilla, H. Ochoa, F. Guinea, Phys. Rev. Lett. 113, 077201 (2014)

\bibitem{QT2}
H. Ochoa, F. Finocchiaro, F. Guinea, V.I. Fal'ko, Phys. Rev. B 90, 235429 (2014)

\bibitem{Edmonds} 
A.R. Edmonds, {\em Angular Momentum in Quantum Mechanics}, 2nd edition, Princeton University Press (1960). 

\bibitem{FootnoteFig} 
For several of the lowest eigenvalues, the inclusion of higher-$l$ spherical harmonics leads to a $<3\%$ change at the distances shown. 

\bibitem{FootnoteA} 
For a trion, $\theta_1=\theta_2$, and the iterative procedure can be used separately 
for symmetric/antisymmetric ($s/a$) states. 

\bibitem{FoootnoteC} Here, we omit the superscript $(1,2)$ as $\gamma_{1}=\gamma_{2}$ for the symmetric
case. Otherwise, the strongest potential minimum has to be chosen.

\bibitem{FootnoteB} 
The small values of the $X^{D/A}$ binding energies also agree with the analytical solution
of Eq.~\rref{finalmatching} obtained for asymptotically shallow ground states in the potential $-\gamma^2/r^2$:   
$\tilde{\epsilon} \sim \exp\left(-\frac{\pi-2{\rm Arg}\left[\Gamma(i\gamma_{1,2})\right]}{\gamma_{1,2}}\right)$ , 
where $\Gamma$ is the gamma function. 

\bibitem{Zhu}
C.R. Zhu, K. Zhang, M. Glazov, B. Urbaszek, T. Armand, Z.W. Ji, B.L. Liu, and X. Marie, Phys. Rev. B {\bf 90}, 161302 (2014).

\bibitem{HeinzU}
T. Heinz, private communication.

\bibitem{PotemskiU}
M. Potemski, private communication.

\end{thebibliography}
\end{document}